# IT Managers' Perception and Response to Digital Disruption: An Exploratory Study

**Alemayehu Molla**
School of Business IT and Logistics
RMIT University
Melbourne, Australia
Email: alemayehu.molla@rmit.edu.au

**Vanessa Cooper**
School of Business IT and Logistics
RMIT University
Melbourne, Australia
Email: vanessa.cooper@rmit.edu.au

**Vass Karpathiou**
School of Business IT and Logistics
RMIT University
Melbourne, Australia
Email: vass.karpathiou@rmit.edu.au

## Abstract

Digital disruption introduces technology-enabled changes at a pace and scale that fundamentally shifts established ways of life and work. This paper addresses the ever changing world of digital disruption but more importantly how it affects the IT people that design and manage the technological changes that lead to digital disruption. The paper draws on i) situational awareness theory to identify IT managers' perceptions of digital disruption, and ii) disruptive innovation theory to identify IT managers' responses to digital disruption. A case study of senior IT managers in an Australian university identifies that IT managers perceive digital disruption from the technological, learning and sense-making perspectives. IT managers' first-order responses to digital disruption were shown to address resource, process and value-based issues while second-order responses focused on the need to build dynamic capabilities in order to be prepared for future digital disruption.

## 1   Introduction

Digital disruption refers to fast transformation of established ways of value creation, social interactions, doing business and more generally our thinking by digitally-enabled mechanisms, or digital technologies (Karim and Walter 2015). It is about innovating to break the status quo, redefine the norm, and change markets and competitors (Weill and Woerner 2015). Digital disruption occurs from individual through industry to societal levels (Deloitte 2012; Reimer et al 2015). For example, at the individual level, mobile connectivity and social media disrupt established work-life boundaries and lead to work-life blur. For business practices, social media, the Internet of Things and real time analytics track interaction, create real-time and personalised experience analysis, and make use of them for targeted end-users. For industry structures, user-generated digital content is changing the ways traditional content production channels operate. At societal level, digital disruption is transforming the traditional practices of public opinion making (Reimer et al 2015).

Existing research on digital disruption addresses the issues of connectivity (Danneels 2004), diffusion of power from established institutions to consumers that challenge institutions' authority (Schmidt and Cohen 2010); technological enablers and affordances that create digital disruption (Tan et al. 2015); and the impact of digital disruption on existing business models (Weill and Woerner 2015). Others have looked at how m-commerce and digital technologies, primarily found in platforms such as eBay and Uber, build strategies and create value by enabling interaction among end-users (Gawer and Cusumano 2010). A much less researched and discussed area, but one that is important to understanding the management and value of digital disruption, is IT managers' understanding and response to this emerging phenomenon (Deloitte 2012; Foster 2014).

IT departments have had to respond to disruption over an extensive period of time. Although the term disruption has not been incorporated into business diction until recently, disruption has been taking place since the development of technology and its mass adoption in industry (Christensen 2000). The areas where this disruption has manifested itself is in the 'constant change' that is prominent in IT delivery and various organisational developments. IT departments have been involved in an innovation role and have served as a centre of innovation (e.g. IBM) (Beath et al. 2012). Some have taken a financial aspect and become cost centres; others have adjusted their objectives and evolved as either service and/or value centres (Beath et al. 2012). Naturally IT departments are expected to both architect and manage the impacts of digital disruption in a sustainable way. However, IT managers unknowingly can create barriers to digital disruption if strategies applied undermine the existing corporate values. It is important to understand how the pressure of change affects IT managers, the work environment, and what needs to be done to mitigate the challenges and capture the potential for these changes so IT departments are responsive with a strong digital presence.

The main questions that we set out to explore are (a) how do IT managers perceive digital disruption and (b) how do they respond (or intend to respond) to digital disruption? These questions were addressed via exploratory interviews with senior IT managers from a leading Australian university's IT department. We have chosen a university case for two reasons. First, increasing cost pressures, government deregulation, and massive open online courses (MOOCs) models are disrupting the way universities are delivering courses. Students are consuming educational content on a global scale. With increasing demand and expectation for student digital experience, IT in universities is not insignificant and in most cases represents the second largest expenditure next to property. These trends and innovations require IT managers in the education sector to change perceptions and to become active architects of implementing new education and business models. Second, a report from Deloitte (2012) forecasted that higher education is one of the sectors that will experience significant digital disruption within the next five years. The lessons from the case could provide important insights to the preparedness of the sector to take advantage of digital disruption opportunities.

## 2    CONCEPTUAL BACKGROUND

### 2.1   Awareness of Digital Disruption

In order to conceptually anchor the exploration of IT managers' perception of digital disruption, we draw from situational awareness theory. Situation awareness is concerned with goals and attention for acquiring and interpreting information from the environment to form mental models and goal-directed behaviour (Endsley 1995). At a very basic level, situational awareness refers to "appropriate awareness of a situation" (Smith and Hancock 1995). In a more comprehensive definition, situational awareness implies "the perception of the elements in the environment within a volume of time and space, the comprehension of their meaning, and the projection of their status in the near future" (Endsley 1995). The three key concepts of situational awareness that are relevant to this study are perception, comprehension and projection.

Based on the situation awareness concepts, we were *first* interested in the IT managers' perception of digital disruption. Digital disruption is a complex phenomenon and subject to many interpretations and mental models. Two of the most common mental models of digital disruption are digital disruption as (a) technological disruption and (b) as sense-making mechanism for transformations in social, business and individual practices due to the convergence of digital and non-digital technologies. As a technological disruption, the law of disruption highlights a simple but an unavoidable principle of the modern life as "technology changes exponentially, but social, economic, and legal systems change incrementally" (Downes 2009). This gap is now getting wider and humans are struggling to keep up with the technological pace (Downes 2009). Marshall (2014) identified social media explosion, mobile revolution, analytics, and cloud enablement as observable elements of digital disruption. Social online networks with their collaboration and information sharing capabilities are now evolving as new business models for value creation. Increased use of geographical-specific services is creating new capabilities for the mobile users through global positioning and target retail systems. Advanced analytics help organisations to get deeper business intelligence and market insights. Various cloud computing models are also enabling new business models, disrupting existing business models with strict institutional networks.

On the other hand, others contend that it is important to understand that digital disruption does not merely mean digitisation of an existing business model or its replacement with a digital alternative (Karim and Walter 2015). In this perspective, digital disruption refers to our understanding of what counts as meaningful, valuable and the right way of performing business practices, giving meaning to, or understanding of, our business and work-life practices (Riemer et al 2015) and how people within that culture socially respond and innovate for themselves (Lindgren 2013). Foster (2014, p 54) quoted from an 'Information Industry Outlook' report where it was stated that, "we see information and technology converging, with devices, humans, value chains, content, and workflows blending to become one. It's a metatheme we call 'Convergence Now!' and it affects virtually every aspect of work and life and will greatly influence innovation. Its impact is wide-ranging, particularly in emerging areas such as health IT, educational technology, marketing services, and marketing automation". This implies that digital disruption is one way of making sense of the changes happening around us.

***Second***, aligned with the concepts of situational awareness, we were also interested in IT managers' comprehension and projection of digital disruption. Digital disruption is "neutral…[and] as with any kind of change, it brings opportunities as well as challenges " (Deloitte 2012). This means it can be comprehended and its impact managed both as a threat (negative) and an opportunity (positive) in a given context. In terms of potential digital disruption threats, rapid changes in digital technologies are impacting existing business practices by invalidating it. As a result, board members estimated that 32% of their revenue would be under threat from digital disruption (Weill and Woerner 2015). Other potential threats of digital disruption include pressure on prices and margins, unexpected source of competition, winner-takes all dynamic and growing talent mismatch (Hirt and Willmott 2014),

On the other hand, digital disruption also offers new opportunities for the creation of innovative business model to compete with established businesses (Hirt and Willmott 2014). Digital disruption can facilitate better product experiences, building stronger customer relationships, and bringing it all to market faster (McQuivey 2013). It enables new business or operating models, such as peer-to-peer product innovation or customer service. Management decisions could also be improved as algorithms crunch big data from social technologies or the Internet of Things. According to Deloitte (2012) "understanding and awareness can help businesses fend off disruptions or indeed take advantage themselves". The next section lays the foundation for responses to digital disruption.

## 2.2  Response to Digital Disruption

The conceptual foundation for responding to digital disruption is anchored on disruptive innovation theory which explains why companies succeed or fail to respond to disruptive innovations (Christensen and Raynor 2003). Our application of the theory, however, is particularly focused on the resources, processes and values (RPV) concepts (Christensen and Raynor 2003; Karimi and Walter 2015). The RPV helps managers of organisations to evaluate the capabilities and weaknesses in responding to digital innovation (Karimi and Walter 2015). Resources include users, equipment, technologies, product designs, funds and relationships. Processes are the procedures and operational patterns of an organisation, whereas values are the standards users use to set priorities for decision-making. Practising these values constitutes the culture of any organisation or businesses (Lucas and Goh 2009). Organisational culture helps to determine what organisations do and what they cannot do, which either facilitate or inhibit the adoption of digital innovation. Managers who wish to prepare for digital changes should carefully revisit their RPV capabilities. Karimi and Walter (2015) have studied the need for capabilities to be dynamic to respond to the dynamic environment the newspaper industry is facing. Their empirical results report "…dynamic capabilities that are created by changing, extending, or adopting a firm's existing resources, processes, and values are positively associated with building digital platform capabilities, and that these capabilities impact the performance of response to digital disruption". Further, Karimi and Walter (2015) assert that IS Managers focus in "building dynamic digital platform capabilities and reinventing their core functions to accelerate digitization". Strategic capabilities and the overall strategic direction of organisations seem to be the current direction of enterprises (Kane et al. 2015). These capabilities can be strengthened by 'experimenting' and using digital technologies that are outside of their current, established business structures (Riemer et al. 2015).

In the education industry, many predict that recent innovations— including low-cost online universities, competency-based instruction, online partnerships between for-profits and traditional universities, and MOOCs— will quite literally transform higher education, as they threaten the future of large numbers of traditional educational institutions (Horn 2013). As an example, Stanford, MIT, Harvard are now conducting some of their classes online to give worldwide students the opportunity to education never before within their approach (Hyman 2012). A number of other universities are offering MOOCs via the Courseera platform and others are mobilising their learning capabilities by harnessing the potential of social software tools (Cochrane 2009). Although there are certainly political and policy obstacles to creating online educational opportunities in higher education, a great deal of innovation can take place outside of the reach of regulation (Horn 2013).

# 3 RESEARCH METHOD

For this study, we have adopted the case study method to qualitatively analyse, explore and capture IT manager's perception and response to digital disruption (Yin 2014). The case study method provides a grounded approach for understanding the two "how" research questions (Yin 2014). We selected a university and asked IT managers to share their perceptions of digital disruption and to describe how they respond (or intend to respond) to digital disruption. The university reported to be experiencing digital disruption in the form of increased competition through the availability of online education courses, including MOOCs. Further, student expectations for IT-enabled services at the university have been increasing and the IT department needs to stay ahead of these rising expectations.

The university is one of the largest universities in Australia based on student numbers, offering vocational education and training as well as higher education study programs. The university strives to be a "global university" with three campuses located in Australia, three overseas and a strong presence in the Asia-pacific region thorough significant off-shore partners. The IT department is one of the six resource portfolios under the university's organisational structure. The primary role of the IT department is to provide the university with information and communication technology in support of research, learning and teaching and administrative activities. The department employs approximately 500 IS staff, 300 full time and 200 contractors. We negotiated access (Pan and Tan 2011) to individually interview the CIO and six directors of the IT department (see Table 1). For the purpose of our research questions, these participants were appropriate as they constitute the senior IT leadership in the university, each having at least 15 years' experience in the IT industry.

|  | Role | Profile |
| --- | --- | --- |
| CIO | Leads the IT department and responsible for: IT planning and strategy for the university and managing technology and information resources. | A track record in leading the development, deployment and operation of large-scale technology systems in a range of sectors such as banking, logistics, defence and higher education. |
| Director of Client Computing | Manages all aspects of IT support. | Has worked in technical support, architecture and operations management. |
| Director of IT Applications | Develops, supports and maintains applications across the university | A track record of IT roles focused on general management, consulting and service delivery in a range of sectors |
| Director of IT Enterprise Architecture | Plans, designs and governs university IT systems. | Experience in web development, management, learning systems and architecture. |
| Director of IT Service Management | Manages day-to-day operation of the IT business unit. | Specialising in organisational change, strategy development and implementation, continual process improvement, management of university applications. |
| Director of IT Engagement and Delivery | Engages with senior management and implements IT solutions to support the university strategy | Experience in architecture, specialising in networking, storage and dat centres; with a track-record in leading IT teams and projects. |
| Director of Infrastructure Delivery | Manages the operation, development and maintenance of computer systems, networks and databases | Business expertise in management, budgeting and forecasting, strategic planning; technical expertise in networking, security, physical and virtual compute and data centre management |

*Table 1: List of interview participants*

The interviews were open ended and focused on eliciting participants' perceptions and experience of digital disruption, generally, and in the higher education sector and their university in particular. Questions also explored how the IT department and the university responded (or intends to respond) to digital disruption. The interviews lasted between 60-90 minutes each, were audio recorded and transcribed. The data analysis was guided by the conceptual foundation and focused on perceptions of digital disruption, comprehension and projection in terms of the opportunities and challenges this brings; and responses to digital disruption.

## 4 FINDINGS AND DISCUSSION

The main findings from the exploratory study are summarised in Table 2 and discussed in the ensuing section.

| Area | Main Findings |
| --- | --- |
| Perception of digital disruption | Technology disruption<br>Learning model disruption<br>Sense-making of changes in the higher education sector |
| Comprehension and projection of digital disruption | Opportunities<br>• Enriching students' experience<br>• Mitigating regulatory and funding pressure<br>• Rethinking the provision of technology<br>Challenges<br>• Proliferation of online teaching and learning models<br>• IT departments' preparedness<br>• Redefinition of value<br>• Transformation of IT roles |
| First order response to digital disruption (resources, processes, values) | Building digital presence<br>Consumerisation of IT<br>Transforming software development models<br>Redefining IT hardware provisioning<br>IT services alignment<br>Institutional role of IT |
| Second order response to digital disruption (dynamic capabilities) | Unmatched value proposition<br>New pedagogy models<br>New connections<br>Chief digital office |

*Table 2: Key findings*

### 4.1 Digital Disruption Perception

The interview participants reinforce the view that digital disruption is not a single phenomenon (Table 3). From a technological perspective, digital disruption is perceived as the entrance of new technology in the market that is disrupting the established business models. From a learning perspective, MOOCs were considered as symbolic examples of digital disruption in the higher education sector. From a sense-making perspective, digital disruption is perceived as the convergence of different changes that is challenging the core of the university sector and the learning pedagogy.

| Perception | Interview log (source) |
|---|---|
| Technology disruption | *"The ERP providers…made their profits from selling software and implementations and you running all the kit yourself [in your organisational environment]. That model is now being disrupted by cloud providers, and that's going to have an impact on us…" (CIO)* |
| | *"Students still want to come on campus, but they want to be connected anywhere they are on campus…bring your own device is another disruption… that has already arrived…" (Director of IT Enterprise Architecture)* |
| Learning model disruption | *"…we see it very much as affecting how things are taught and learned. Obviously the tools and the skills to do that. We see it as relating to the great spectre of MOOCs and all things that might compete with the university in the fabric of what it is" (Director of IT Applications)* |
| Sense-making | *"Digital disruption in our sector is much more about the global learning and the MOOCs. It won't be MOOCs that disrupts us, it will be the next thing. It will be the thing that we have learned from the MOOCs and then we've improved…MOOCs is a massive experiment, and we have learned truck load as an industry and as a global sector that's going to be in the news…" (CIO)* |

*Table 3: Perception of digital disruption*

## 4.2 Opportunities of Digital Disruption

Digital disruption offers universities opportunities for enriching students' experience through blended educational delivery, mitigating regulatory and funding pressure, and rethinking technology provision.

Educational institutes in general are now exploring different ways to incorporate new business models, such as MOOCs as part of their existing business processes and practices. As stated by the CIO, there are fairly high levels of expectations within the institute towards digital disruption. There is a need to understand what role the university is playing in terms of the root desire of the customers and to redefine the business model to offer rich student experience. The re-definition of the pedagogy model may help the university to face the threat of converging global supply and demand as highlighted by Hirt and Willmott (2014). This was seen as something that is currently growing in some business areas, but eventually becoming a reality across the board, for example:

> *"One of the things in digital age is you don't talk about giving people choice. You talk about helping them choose…that is where we would assume that people will come to us…" (Director of IT Applications).*

> *"Global learning by design is the key project that I think is really focused on the delivery side of digital disruption." (CIO)*

The higher education sector is faced with a number of political and policy constraints and rising up pressure on prices and profit. Digital disruption can be used to drive a great deal of innovation outside the reach of the regulation (Horn 2013).

> *"…we have always had it in our mind that the universities do need to change. We have obviously got a lot of threats from the government as well from a financial point of view. We have got to figure out ways to effectively deliver what we need to deliver and doing it online is one way of doing that…"* (Director of Client Computing)

In term of technology provision, digital disruption is seen as a change agent for different service provided by the university in terms of collaboration tools such as various Google apps. It is also transforming application delivery to be more customer-centric, deliver value faster, and react more quickly around the disruption.

> *"… I mean, we need to change our application mix because we need to be more customer-centric, we need to actually move ourselves into a digital age and that in itself will change our mix, the same, designing for mobile first. Success factors will run on a tablet but you wouldn't want to be doing your work patterns on a tablet using what we have at the moment." (Director of IT Applications)*

## 4.3    Challenges of Digital Disruption

Some of the challenges of digital disruption identified in the case include the proliferation of online teaching and learning models, IT departments' preparedness, redefinition of value, and the transformation of IT roles.

Online teaching and learning models, such as, OpenCourseWare, MOOCs, TED, Wikiversity are seen as threats to the future of traditional educational institutions. Although some considered this as hype and an overreaction, the momentum and potential behind online model are forcing established institutes to evolve their pedagogical models and associated technology service.

IT changes are caused both when IT systems fail to provide business utility as well as when IT vendors phase out technologies and make them unsupportable. IT departments in universities and other similar types of organisations need to balance these two forces.  University IT departments are externally facing with more contact points with vendors and technology leaders. As a result, they can be more receptive to digital disruption. However, their effort to promote change within the IT department and to push along projects in the wider institution are constrained by the business units as well as IT's expectations of the business unit to make the first move.

> *"…there is a reticence for big things to be owned out in the business we find because I guess people are worried about the politics and could they be blamed for something if it all went pear shaped…We couldn't lead the revolution there…We could definitely support it, promote it and help a revolution but it needs to start with academics and the academic community wanting to change and asking for our support to do that…"* (Director of IT Engagement and Delivery).

Digital disruption such as MOOCs challenge the value expectations of students in terms of 1) the features they want in the course, for example, student will not simply select a course because it is cheap, 2) the sign value, which refers to the institutional brand, and 3) the economic value.

> *"As you move into more of an experienced economy that shifts, in my belief, much more towards sign value rather than feature value…I think we are far behind in feature value…in some respects we are also behind the game in sign value…"* (Director of IT Applications)

The impact of digital disruption is not only disrupting the educational business models but also challenging the IT roles associated with educational business model.

> *"…I would expect that [role of IT in the management of network security] to go out completely.  I would expect identity to be completely out. We are already seeing massive migrations to Facebook as an identity platform and Google as an identity platform. You are seeing that on applications like LinkedIn and all of those popular apps are moving…I think over the next five or ten years you will start to see a real – a genuine dissolving of corporate infrastructure as you know it…You start to see infrastructure as software everywhere…"* (Director of Infrastructure Delivery).

## 4.4    First order response to digital disruption

The case study indicated that IT managers can respond to digital disruption through a first order response of resources, processes and values. This involves the reconfiguration of their resources (what they have), improvisation of how they do things (processes) and redefining what IT departments and the core business they service stand for (values).

**In terms of resources**, one of the main responses to digital disruption that is shared by the IT managers is the importance of building digital presence. For the next five years, digital presence through the use of web and mobile technologies to position the institute's brand and in the delivery of products, were seen as key priorities. These priorities can provide good value proposition in the market to drive efficient delivery cost base, and to scale up students' capabilities. In the words of one interviewee,

> *"…we need to really get inside what's the reason that people buy, because they don't buy … because it is a thousand bucks less than … they're buying into a community… if you can't present a digital presence that resonates with that community, then I think that you have lost the argument…"* (Director of IT Applications)

Consumerisation of IT by revamping traditional platforms with Bring Your Own Devices (BYOD), Bring Your Own Cloud (BYOC), and Bring Your Own Network (BYON) models are now making their way into the educational environments to enable students to work in their preferred ways,

> *"...we are attempting to ensure that the students can use the equipment that they have and bring it and it will work. So services like myDesktop are a way to attempt to ensure we can bring a defined environment to the students that they can use as part of their studies but they can choose to use it on their tablet or on their laptop or even on their phone..."* (Director of IT Enterprise Architecture)

**In terms of process responses**, the IT managers have identified the need for transforming software development models from traditional plan-driven models to agile methodologies in order to be able to deliver value faster and react more quickly around the disruption.

> *"...we're deploying quite heavily now agile delivery methodologies to make that more responsive and more sustainable...I want to make sure someone who follows me up in ITS is really working out how to take the ITS team on that agile journey; continue that journey..."* (CIO)

Further, IT departments are also changing the traditional ways of IT hardware provision both to their internal users as well as their clients (in this case students) that require access to hardware resources:

> *"...except for the ones that want to hard compute, they want CPU cycles or whatever – people don't want a computer to have a computer, they want a computer because they need to do something on it...so our overall philosophy on the client computing space is the applications that people need should be associated with them, not with the device that they happen to be on at that time..."* (Director of Client Computing)

IT departments need also to align IT services so that they reflect the demands and expectations of the digital age which is both about giving people choice as well as helping them to choose. In the case study, different IT teams coordinate with different operational departments without necessarily aligning their internal IT processes and practices. For example, the web services group is closely linked to marketing; the network administrative group facilitates operations of student access to computers, labs, and other technical services. Teaching and learning tools such as, Blackboard are under the realm of the Office of Dean, Learning and Teaching facilitated by the IT department. Addressing this to provide uniform experience to students required a procedural change.

> *"I hope, [we] will bring this alignment [within the IT department] because it is very hard when you get into the broad academic space...with so many different disciplines that are trying to do so many different things... "* (Director of IT Enterprise Architecture)

As regards **value response**, there is a need to redefine the institutional role of IT as a leader of change in addition to as supporting and enabling business unit changes:

> *"...we feel like we've already tackled many of those frontiers and the uncomfortable edge of leading business change when actually the business should be leading business change. That's one of those areas where we have got 'That is too big for us'."* (Director of IT Engagement and Delivery)

Another interviewee reinforced the need for redefining what IT departments stand for as follows:

> *"...So are we ready for it? Yes. Do we know exactly how we will answer it? No, it depends how they want to answer MOOCs. Like for example there are a couple of different types and styles of MOOCs. So it depends on what style is happening. Are we ready for it emotionally? Yes. Do we have the technology foundation for it? Yes..."* (Director of IT Infrastructure Delivery)

### 4.5　Second order response to digital disruption

In addition to first order resources, processes and values, digital disruption response requires creating dynamic capabilities and building the ability to utilise the RPV components to enable the organisation to react and adapt faster to changes.

In order to prepare the institute to compete against new digital entrants, the university took a strategic decision aligning existing organisational assets and capabilities with the new global learning business models. The aim and objective of this decision was to use the existing physical infrastructure to create not only a good online experience, but also a good classroom experience.

> *"We have got an offer that other online peer player can't match, because they don't have the physical infrastructure. The barrier to them in that space is you can't build it, there's a big capital investment...the value is...that you can come in and play with a laser, you can come and play with the largest 3D printer in the Southern Hemisphere...can come and physically get experience in a wind tunnel-- how are you going to do an online wind tunnel? You're not."* (CIO)

Further, executive decisions to use existing institute resources and capabilities to develop new pedagogical models to deliver MOOC-like services are typical higher order responses to digital disruption,

> *"...we invented a thing called MOOCENSTEIN which was our concept of...we've pretty much politely been asked to do nothing... what I mean by that is, don't go out and overtly create some digital strategy to go and rule the world and change the way the pedagogy works...so we've played a very much 'Okay we won't'...Now that is starting to come about as a result of the global learning by design strategy..."* (Director of IT Engagement and Delivery)

In addition, higher order dynamic capabilities would require going beyond forcing learning tools such as Blackboard and to experiment by connecting students' preferred networks with institutes systems.

> *"...I think the global learning by design is about how do we leverage and learn from the digital disruption that is happening and bring those bits that fit our strategy, how do we change our strategy where it need to be tweaked."* (CIO)

The appointment of Chief Digital Officer to orchestrate response to digital disruption would also be required in building dynamic capability. This is consistent with some universities experience that have transformed the role of *Chief Information Officer* to *Chief Digital Officer* who will be concerned less about running IT or provisioning of IT and more about advising how to operate a university in the digital economy (Foster 2014).

> *"...how you integrate cloud solutions and actually bring that together, they need to be a chief advisor there...how technology is leveraged to make the university successful to achieve the university strategy...less you concern yourself with the lower end of the value chain...the more you can focus on the strategy and using that to enable the strategy"* (CIO)

# 5 Conclusion

In the present IT environment, flexibility and agility have been framed to reflect the need to respond to digital disruption. This requires not only understanding disruptive technology but also how to transform and build organisations in a digital economy and on a digital scale. In this paper we set out to answer two research questions, that is, (a) how do IT managers perceive digital disruption? and (b) how do they respond (or intend to respond) to digital disruption? With respect to the first research question, we have drawn from situational awareness theory and identified that IT mangers perceive digital disruption from the technological, learning and sense-making perspectives. Further, IT managers recognise a number of opportunities associated with digital disruption, including the potential to enrich the student experience; mitigate regulatory and financial pressures; and rethink the provision of technology. Yet IT managers also perceive a number of challenges associated with digital disruption. These include the proliferation of online teaching and learning; the preparedness of the IT department to cope with the digital disruption; the redefinition of value; and transformation of IT roles. With respect to the second research question we have drawn from disruptive innovation theory and shown that IT managers' response to digital disruption comprise first-order and second-order responses. First order responses address resource (e.g. building a digital presence, consumerisation of IT), process (e.g. transforming software development models, redefining IT hardware provision and IT service alignment) and values (e.g. institutional role of IT). Second-order responses focus on the need to build dynamic capabilities (e.g. unmatched value propositions, new pedagogical models, new connections and chief digital office) in order to be prepared for future digital disruption.

In addressing these two research questions we have contributed to the limited research on digital disruption from the perspective of IT managers in organisations via an exploratory case study of senior IT managers in an Australian university. The research has a number of limitations, however, that offer the potential for future research. First, we conducted a single exploratory case study of an Australian university. Future studies might usefully conduct multiple case studies on digital disruption in other sectors. While our qualitative approach has provided insight into digital disruption within its real-life context, quantitative studies would enable a wider set of managers to be investigated and thus generalisation of the results.

## Acknowledgements

This research was made possible through RMIT's Sustainable Urban Precinct Program grants.

## Copyright